\documentclass[aps,preprint,prd,epsfig,axodraw,nofootinbib]{revtex4-1}
\usepackage{amssymb}
\usepackage{amsmath}
\usepackage{graphicx}
\usepackage{fancyhdr}
\usepackage{epsfig}
\def\br{\begin{eqnarray}}
\def\er{\end{eqnarray}}
\def\be{\begin{equation}}
\def\ee{\end{equation}}

\def\({\left(}
\def\){\right)}

\def\lesssim{\mathrel{\hbox{\rlap{\hbox{\lower4pt\hbox{$\sim$}}}\hbox{$<$}}}}
\def\gtrsim{\mathrel{\hbox{\rlap{\hbox{\lower4pt\hbox{$\sim$}}}\hbox{$>$}}}}
\begin{document}

\title{Fixing the number of non-sequential generations within the $SU(2)_{L}\otimes U(1)_{Y}$ gauge group}
%\end{center}

\author{Elmer Ramirez Barreto\footnote{elmerraba@gmail.com} and David Romero Abad\footnote{dromero@usil.edu.pe}}
\affiliation{{ Grupo de Investigaci\'on en F\'isica, Universidad San Ignacio de Loyola\\ Av. La Fontana 550,
	La Molina, Lima, Per\'u.}}

%\author{Elmer Ramirez Barreto}
%\affiliation{Grupo de Investigaci\'on en F\'isica, Universidad San Ignacio de Loyola\\ Av. La Fontana 550,
%	La Molina, Lima, Per\'u.}

%\author{David Romero Abad}
%\affiliation{Grupo de Investigaci\'on en F\'isica, Universidad San Ignacio de Loyola\\ Av. La Fontana 550,
%	La Molina, Lima, Per\'u.}
%elmerraba@gmail.com
%dromero@usil.edu.pe

\begin{abstract}
In this work we explore the possibility to fix the number of new non-sequential chiral-type generations of fermions that
could be  added to the standard model  by combining the condition that arise from the anomalies cancellation with the 
restriction in the number of flavors impose by the QCD asymptotic freedom.  We found that the maximum number of 
new  generations is four, and that allows us at the same time, to place limits for the electrical charges of quarks and leptons within
a SM-like framework.
Our  result is compatible with the constraints  involving
new generations and their contributions to the oblique parameters $S$ and $T$. 
%\\
%PACS: 14.60.St; 14.60.Pq; 12.60.Cn; 12.60.Fr.
%

%\keywords{QCD asymptotic freedom; New fermions; Gauge anomalies cancellation.}
\end{abstract}

\maketitle

\section{Introduction}	

Although the Standard Model of strong and electroweak interactions \cite{SM1,SM2,SM3} (SM)  is in agreement with all experimental measurements, 
we have reason to believe that it is a low-energy effective theory 
of an ultimate fundamental theory of nature, 
able to answer many of the
open questions in the SM, such as the neutrino masses, the hierarchy
problem, the number of generations; matter-antimatter asymmetry of the Universe;
the strong CP problem and the dark matter content of the Universe  among others. 

Concerning the number of generations, the SM offers neither an explanation or a justification for why we have three generations
of quarks and leptons. Experimentally, we know, from the $Z$ invisible decay, that the number of light neutrinos is equal to three \cite{Z3}, 
which implies a  compelling proof that there are only three conventional neutrinos with mass below $M_{Z} /2 \simeq 45$ GeV, and, 
by extrapolation, it leads to the idea that there are only three quark-lepton families.
But, what is the principle limiting the number of chiral families? Why not have a fourth generation or even more? 
 or why not have leptons or quarks with exotic electric charges?

From the theoretical side, the only upper limit comming from the QCD asymptotic freedom allowing us to include at most sixteen quark flavors or eight
quark families \cite{QCD1,QCD2}. 
On the other hand, the chiral anomalies cancellation require to have the same number of quark and lepton families, being simple 
replicas of the first families:
$(\nu_{e} \ e^{-})^{T}_{L}$ and $(u \ d)^{T}_{L}$. Although the discovery of the Higgs boson has strongly restricted the possibility 
of having a sequential fourth family, recent works have shown that by extending the scalar sector, like the 
Type-II two Higgs
doublet model (2HDM) \cite{2HDM}
and making use of  the so called exact wrong-sign limit where all of the down-type fermions have
opposite Higgs coupling to the up-type fermion \cite{FERRE},
a fourth family $(t' \ b')^{T}_{L}$
is still allowed without altering the Higgs production \cite{Higgs4f}. 
It is possible because the contributions of the  new heavy quarks to the Higgs production via gluon fusion and the
respective contributions of the new heavy leptons (necessary for anomalies cancellation) to the  $\gamma\, \gamma$ and $Z\,\gamma$ Higgs decay channels
 may cancel each other and so, there is no enhancement in Higgs production through
gluon-gluon fusion. However, from the heavy Higgs phenomenology, 
this approach seems to be useful when higher order corrections are considered \cite{SIN}.

\newpage
Moreover, as it is known, the presence of new fermions can affect the so-called oblique parameters $S$ and $T$ which represent
the effects of new physics on the $W$ and $Z$ vacuum polarization amplitudes (both parameters describe weak-isospin-symmetric
and weak isotriplet contributions to $W$  and $Z$ loop diagrams, respectively) \cite{PESKIN}, and whose limits are well established
through the electroweak precision data \cite{PDG}.  Thus, by using these parameters, very interesting possibilities have been found 
regarding the inclusion of new generations of quarks and leptons within or outside the SM framework. In this sense, some authors have
found possibilities to have up to two new generations if the neutrino is Dirac-type and up to five new generations if the neutrino is Majorana-type,
 demanding small differences in mass within the new $SU(2)$ multiplets \cite{STU, TAKEO, Extra, ADRIANO}. If these mass splittings
are not small, the compatibility between  new generations and the $S$ and $T$ parameters is lose. 
To save this situation we must include new physics contributions, such as new scalar doublets, for example \cite{HE}.

Otherwise, given the experimental interest of the LHC in the search for quarks and leptons with 
exotic electrical charges  $(5/3\:e, -4/3\: e, -2e,\: \text{etc.})$, \cite{newLHC}
it is possible to include, for example, $SU(2)_L$ chiral doublets  beyond the third generation in a nontrivial way, 
with different hypercharge from the SM replicas (containing quarks and leptons with exotic electric 
charges), keeping it free of anomalies and without the need to extend the SM gauge group or go to
grand unified theories (GUT) \cite{ALVES, BIZOT, FRAMPTON}. 
On the other hand, the inclusion of new chiral
fermions in higher representations of $SU(2)_L$ have been studied in
the context of the unification of the gauge couplings at a high energy scale \cite{CEBOLLA,SIMOES}.

In this letter, we want to show that by including quarks and leptons  with exotic charges in the fundamental representation of the SM gauge
group, we can fix the number of  doublets allowed within the SM, if we combine the cancellation of the anomalies and the  asymptotic freedom of
the QCD. Additionally, by setting the number of new non-sequential generations, a limit may be placed for exotic electrical charges for both new quarks and
leptons.

So, in the next section we will show how to include new quarks and leptons with exotic charges in a safe way, starting from the 
conventional electric charge operator and analyzing the cancellation of the chiral  anomalies arriving
to  the result of being able to have a limited number of new non sequential generations and 
and allowing us to place constraints to the values ​​of the electrical charges of both leptons and quarks.
Finally we will present  our conclusions  and final remarks.

\section{New non-sequential generations in a SM-like framework}

The inclusion of sequential families in the SM  is straightforward if we consider the  hypercharge values $\mathcal{Y}_{SM} = 1/6$ (quarks),
and $\mathcal{Y}_{SM} = -1/2$ (leptons) for these new generations.
By considering the electric charge operator:
\begin{equation}
 Q/e = I_{3} + \mathcal{Y}
\end{equation}
and the chiral anomalies, which have to be canceled in order to guarantee
the renormalizability of the theory, we have sequential generations 
$(t' \ b')^{T}_{L}$ and $( \nu  \ e')^{T}_{L}$ respectively.
Thus, with this choice for $\mathcal{Y}$ it is clear that we can have 
$n$ additional replicas (called sequential generations)  to the SM generations.

In principle, the hypercharge assignation for the new particles is arbitrary, but as the simplest assumption, we think that the new quarks and leptons will follow the SM structure, integer electric charge for leptons and fractional electrical charge for quarks. So, as it was already shown\cite{ALVES}, the possibility of including quarks with exotic charges (5/3 e , -4/3 e) within the SM gauge group depends basically on 
the increase or decrease of the hypercharge  by one unit with respect to the SM one.
Then, with the new hypercharge $\mathcal{Y} = \mathcal{Y}_{SM} \pm 1 $, and considering the cancellation of the following  anomaly 
equations:

%\begin{widetext}
\begin{eqnarray}
\label{eq:Chiral-anomalies}
\left[\mathrm{\mathrm{SU}(3)}_{C} \right]^{2} \mathrm{\mathrm{U}(1)}_{\mathcal{Y}} \rightarrow & A_{C} &= \sum_{Q}\mathcal{Y}_{Q_{L}} - \sum_{Q}\mathcal{Y}_{Q_{R}}	\nonumber	\\
\left[\mathrm{\mathrm{SU}(2)}_{L} \right]^{2} \mathrm{\mathrm{U}(1)}_{\mathcal{Y}} \rightarrow & A_{L}  &= \sum_{\ell}\mathcal{Y}_{\ell_{L}} + 3\sum_{Q}\mathcal{Y}_{Q_{L}}	\nonumber	\\
\left[\mathrm{\mathrm{U}(1)}_{\mathcal{Y}} \right]^{3} \rightarrow & A_{\mathcal{Y}}&=
	\sum_{\ell, Q}\left[\mathcal{Y}_{\ell_{L}}^{3}+3\mathcal{Y}_{Q_{L}}^{3} \right]	%\\ &&
	- \sum_{\ell, Q}\left[\mathcal{Y}_{\ell_{R}}^{3}+3\mathcal{Y}_{Q_{R}}^{3} \right] 	\nonumber	\\	
\left[\mathrm{Grav} \right]^{2}   \mathrm{\mathrm{U}(1)}_{\mathcal{Y}} \rightarrow & A_{\mathrm{G}}&=
	\sum_{\ell, Q}\left[\mathcal{Y}_{\ell_{L}}+3\mathcal{Y}_{Q_{L}} \right]%\\ &&
	- \sum_{\ell, Q}\left[\mathcal{Y}_{\ell_{R}}+3\mathcal{Y}_{Q_{R}} \right],
\end{eqnarray}
%\end{widetext}
\noindent

where $\mathcal{Y}_{Q_{L}}$, $\mathcal{Y}_{\ell_{L}}$, $\mathcal{Y}_{Q_{R}}$ and  $\mathcal{Y}_{\ell_{R}}$ are the hypercharges for the
$SU(2)_{L}$ doublets and singlets for leptons and quarks, we arrive at the following new doublets that include  quarks 
 $X$ and $Y$ with exotic charges and the 
necessary inclusion of new lepton doublets  
containing leptons  $E'$, $E$ and $F$, plus the corresponding right-handed singlets. 
We will refer to these new fermions as exotic quarks and leptons.

\begin{eqnarray}
\label{qex}
 &  & \mathcal{\psi}_{L}^{X}\equiv\left[\begin{array}{c}
X_{L}\\
U_{L}^{\prime}
\end{array}\right]\sim\left(\mathbf{2,\,}7/6\right),\quad X_{R}\sim\left(\mathbf{\,1,\,}5/3\right),\quad U_{R}^{\prime}\sim\left(\mathbf{1,\,}2/3\right),\cr
 &  &\cr
 &  & \psi_{L}^{Y}\equiv\left[\begin{array}{c}
D_{L}^{\prime}\\
Y_{L}
\end{array}\right]\sim\left(\mathbf{2,\,}-5/6\right),\quad D_{R}^{\prime}\sim\left(\mathbf{\,1,\,}-1/3\right),\quad Y_{R}\sim\left(\mathbf{1,\,}-4/3\right),
\end{eqnarray}
and for the leptons
\begin{eqnarray}
 &  & \Psi_{L}^{N}\equiv\left[\begin{array}{c}
E_{L}^{'+}\\
N_{L}
\end{array}\right]\sim\left(\mathbf{2,\,}1/2\right),\quad
E_{R}^{'+}\sim\left(\mathbf{1,\,}1\right),\quad N_{R}\sim\left(\mathbf{1,\,}0\right),\cr
 &  &\cr
 &  & \Psi_{L}^{F}\equiv\left[\begin{array}{c}
E_{L}^{-}\\
F_{L}^{--}
\end{array}\right]\sim\left(\mathbf{2,\,}-3/2\right),\quad E_{R}^{-}\sim\left(\mathbf{1,\,}-1\right),\quad F_{R}^{--}\sim\left(\mathbf{1,\,}-2\right),
 \label{nl2}
\end{eqnarray}
in which the numbers between parenthesis refers to transformation properties under $SU(2)_{L}$ and $U(1)_{\mathcal{Y}}$, respectively.
The singlets $E_{R}^{'+}\sim\left(\mathbf{1,\,}1\right)$,  $E_{R}^{-}\sim\left(\mathbf{1,\,}-1\right)$,  
and $N_{R}\sim\left(\mathbf{1,\,}0\right)$  are irrelevant for canceling the anomalies, once the first two form a vector fermion field, 
and the last has zero hypercharge.
As we see, beside the $X$ quark with electric charge $5/3$, we also have the $Y$ quark  with  electric charge $-4/3$. Thus, the fermion 
content above extend in $\pm 1$ the range of the SM 
particles electric charges allowing for quarks charges $\mp\, 4/3,\,\mp \,1/3,\,\pm \,2/3,\,\pm\, 5/3$,
and for leptons charges $0,\,\pm\, 1,\,\pm 2$.

Now, if we extend our analysis for $\mathcal{Y'}=\mathcal{Y}_{SM}\pm 2$, we must include two new generations of quarks 
containing the additional exotic quarks $Q$ and $\mathcal{Q}$ with electric charges 8/3 e and  -7/3 e  respectively:

\begin{eqnarray}
\label{qex}
 &  & \mathcal{\psi}_{L}^{Q}\equiv\left[\begin{array}{c}
Q_{L}\\
X_{L}^{\prime}
\end{array}\right]\sim\left(\mathbf{2,\,}13/6\right),\quad Q_{R}\sim\left(\mathbf{\,1,\,}8/3\right),\quad X_{R}^{\prime}\sim\left(\mathbf{1,\,}5/3\right),\cr
 &  &\cr
 &  & \psi_{L}^{q}\equiv\left[\begin{array}{c}
Y_{L}^{\prime}\\
\mathcal{Q}_{L}
\end{array}\right]\sim\left(\mathbf{2,\,}-11/6\right),\quad Y_{R}^{\prime}\sim\left(\mathbf{\,1,\,}-4/3\right),\quad \mathcal{Q}_{R}\sim\left(\mathbf{1,\,}-7/3\right),
\end{eqnarray}
 and two new leptonic generations  with the additional exotic leptons $\mathcal{E}^{++}$ and ${F}^{---}$:
\begin{eqnarray}
 &  & \Psi_{L}^{\mathcal{E}}\equiv\left[\begin{array}{c}
\mathcal{E}_{L}^{++}\\ 
E_{L}^{'+}
\end{array}\right]\sim\left(\mathbf{2,\,}3/2\right),\quad
\mathcal{E} \sim\left(\mathbf{1,\,}2\right),\quad E_{R}^{'+}\sim\left(\mathbf{1,\,}1\right),\cr
 &  &\cr
 &  & \Psi_{L}^{\mathcal{F}}\equiv\left[\begin{array}{c}
F_{L}^{--}\\
\mathcal{F}_{L}^{---}
\end{array}\right]\sim\left(\mathbf{2,\,}-5/2\right),\quad F_{R}^{--}\sim\left(\mathbf{1,\,}-2\right),\quad \mathcal{F}_{R}^{---}\sim\left(\mathbf{1,\,}-3\right),
 \label{nl2}
\end{eqnarray}

If we take the new general hypercharge $\mathcal{Y'}=\mathcal{Y}_{SM}\pm n $, with $n = 1,2,3...$, we can include $n_{g} =2\,n$ non-sequential
generations of quarks and leptons, with exotic electric charges expressed in a general form, given by:

\begin{eqnarray}
\label{qex}
 &  & \mathcal{\psi}_{L}^{\mathcal{Q}}\equiv\left[\begin{array}{c}
\mathcal{Q}^{2/3 + n_g}_{L}\\
\mathcal{X'}^{-1/3+n_g}_{L}
\end{array}\right]\sim\left(\mathbf{2,\,}1/6 + n_g \right),\quad \mathcal{Q}_{R}\sim\left(\mathbf{\,1,\,}2/3 + n_g \right),\quad \mathcal{X}_{R}^{\prime}\sim\left(\mathbf{1,\,}-1/3 + n_g \right),\cr
 &  &\cr
 &  & \mathcal{\psi}_{L}^{\mathcal{K}}\equiv\left[\begin{array}{c}
\mathcal{Y'}_{L}^{2/3 - n_g}\\
\mathcal{K}_{L}^{-1/3 - n_g}
\end{array}\right]\sim\left(\mathbf{2,\,}1/6 - n_g \right),\quad \mathcal{Y}_{R}^{\prime}\sim\left(\mathbf{\,1,\,}2/3 - n_g \right),\quad \mathcal{K}_{R} \sim\left(\mathbf{1,\,}-1/3 - n_g \right), \nonumber\\
\end{eqnarray}
and for the leptons
\begin{eqnarray}
 &  & \Psi_{L}^{\mathcal{E}}\equiv\left[\begin{array}{c}
\mathcal{E}^{n_g}\\ 
E_{L}^{-1 + n_g}
\end{array}\right]\sim\left(\mathbf{2,\,}-1/2 + n_g \right),\quad
\mathcal{E} \sim\left(\mathbf{1,\,}n_g\right),\quad E_{R}\sim\left(\mathbf{1,\,}-1 + n_g \right),\cr
 &  &\cr
 &  & \Psi_{L}^{\mathcal{F}}\equiv\left[\begin{array}{c}
F_{L}^{-n_g}\\
\mathcal{F}_{L}^{-1 - n_g }
\end{array}\right]\sim\left(\mathbf{2,\,}-1/2 - n_g \right),\quad F_{R}\sim\left(\mathbf{1,\,}-n_g\right),\quad \mathcal{F}_{R}\sim\left(\mathbf{1,\,}-1-n_g\right),\nonumber\\
 \label{nl2}
\end{eqnarray}

Therefore, the inclusion of new non-sequential chiral-type doublets with exotic electric charges have to be in pairs, i.e.
$n_{g}=2n$  with $n = 1,2,3, \cdots$, in order to maintain the consistency of the model i.e. cancel the gauge anomalies
$A_{C}$=0, $A_{L}$=0, $A_{\mathcal{Y}}$=0 and $A_{G}$=0.

Then, with this general result, a natural question arises, how could we fix the maximum value of $n$? It is clear that if the maximum value for $n$ could be found, 
in addition to setting the maximum number of allowed generations, we can also limit the maximum value of the electric charge that new
leptons and quarks can carry inside this SM-like framework.

So, if we take into account that the asymptotic freedom  of the QCD, tested up to $\vartheta$(TeV) at the LHC \cite{ATLASalpha},
require that  the number of flavors of quarks ($n_{F}$)
 be equal equal to or less than 16, where 6 flavors are assigned to the SM quarks, then we can still 
accommodate 10 new flavors of quarks match up  in 5 new $SU(2)$ doublets. With this information and by considering the relation
$$
n_{g}=2n, 
$$
which implies that the number of new flavors of exotic quarks for a given value of $n$ is equal to 2\,$n_{g}$, we get the
following relation:
$$
2\: n_{g} \leq 10    \,\,\,\,\,\ or  \,\,\,\,\ n \leq \,\,2.5 .
$$

Thus, we have found that the maximum number of new non-sequential generations containing quarks and leptons
with exotic electrical charges that can be included is four. In addition, with the value of $n = 2$, we can limit 
the   allowed electric charges of the new  and SM quarks and leptons:
$$
 - \frac{11}{3} |e| \,\, \leq   \,\, Q_{quarks} \,\, \leq \,\, \frac{14}{3} |e| 
$$
$$
 - 5\,|e| \,\, \leq   \,\, Q_{leptons} \,\, \leq \,\, 4\,|e| 
$$

\section{Conclusions and final remarks}

In this work we have considered the possibility of limiting the maximum number of new  generations that can be included
within the SM framework, but  in one way in which we can have both leptons and quarks with exotic electrical charges. Since the SM does not justify or 
explain why the existence of only three generations and the absence of fermions with exotic electrical charges, the search for a 
solution leads to extensions of the gauge group including  grand unified theories. However, when considering the possibility
of including fermions with exotic electric charges in the SM and maintaining the consistency of the theory through the cancellation of the 
gauge anomalies, we arrive at a scenario where we can fix the maximum number of new doublets that can be added to the SM spectrum, 
and at the same time, include exotic electric charges. 
So, if we consider values ​​of the hypercharge such that the new quarks and leptons accommodated in the new doublets can have exotic 
electric charges, and maintaining the cancellation already mentioned, and if we take into account that the asymptotic freedom
of the QCD also limits the maximum number of flavors of quarks to 16 flavors, we observe that the maximum number of new generations that
can be added to the SM is equal to four, being these new generations of the non-sequential type.
Now, as it has been shown in the literature, the inclusion of new coloured fermions in different $SU(3)_{c}$ representations (triplets,
sextets, octuplets and decuplets) affects the  running of the $SU(3)_{c}$ gauge coupling at high energies,
which could in some cases, cause the loss of QCD asymptotic freedom \cite{ Becciolini, Llorente}. In that sense, our proposal to include new quarks
in the triplet representation of $SU(3)$ allows us to have even  an asymptotic behavior for the  strong gauge coupling.
Besides, with this value for $N_{F}$, we can place constraints for the allowed values ​​of the electric charges for quarks and leptons.

On the other hand, if we take into account that the effect of new generations can generate deviations of the oblique parameters $S$, $T$ 
 very sensitive to new physics, allowing us to limit the number of new generations compatible with the electroweak precision data.
As was already shown \cite{ALVES, HE}, the introduction of new non-sequential generations 
(i.e. with $\mathcal{Y} = \mathcal{Y}_{SM} \pm n $) does not affect the oblique parameters $S$ and $T$, if the splitting between
the masses of the particles in the doublet is small \footnote{The relation
for the $S$ and $T$ parameters is 
show in the appendix}.
Following this reasoning, the proposal of this work to include new doublets is compatible with the  deviations
allowed  for the oblique parameters constrained by the electroweak precision data.

In conclusion, we have shown that by considering the inclusion of new fermions with exotic electrical charges in non-sequential generations
and the maximum number of flavors  allowed by the QCD, it is possible to indicate a maximum number of new non-sequential generations
that can be added to the SM ones. From our analysis we conclude that the maximum 
number of non-sequential generations allowed is four, having at the end a spectrum of seven generations in total within
this SM-like framework,
being compatible with the limits imposed by the QCD and by the electroweak  precision data.  

Finally, we call attention to the fact that the LHC is seeking 
experimental evidence for fermions with exotic charges and masses above electroweak energy scale through the MoEDAL experiment\cite{MoeDal}, and so,
further phenomenological analysis  will deserve our attention in future studies.

\section*{Acknowledgments}

We are grateful to Profs. Alex Gomes Dias and Adriano Natale for reading the manuscript  and for calling the attention to some points.

\section*{Appendix}

In this appendix, we show the  general expressions for S and T ~\cite{HE} involving the contributions of
new fermions $\left(\psi_{1},\psi_{2}\right)$, with masses $\left(m_{1},m_{2}\right)$ and 
whose left-handed components form a doublet $\Psi\equiv\left(\psi_{1L}\,\,\psi_{2L}\right)^T\sim(\mathbf{2,\,}\mathcal{Y})$
of hypercharge $\mathcal{Y}$, and their right-handed components are singlets
$\psi_{1R}$, $\psi_{2R}$,

\begin{align}
S_{\Psi} & =\frac{N_{C\psi}}{6\pi}\left[1-2\mathcal{Y}\,\mathrm{ln}\frac{x_{1}}{x_{2}}+
\frac{1+8\mathcal{Y}}{20x_{1}}+\frac{1-8\mathcal{Y}}{20x_{2}}\right],\label{ps}\\
T_{\Psi} & =\frac{N_{C}}{8\pi s_{W}^{2}c_{W}^{2}}F\left(x_{1},x_{2}\right),\label{pt}\\
U_{\Psi} & =-\frac{N_{C\psi}}{2\pi}\Bigg\{\frac{x_{1}+x_{2}}{2}
-\frac{\left(x_{1}-x_{2}\right)^{2}}{3}+\left[\frac{\left(x_{1}-x_{2}\right)^{3}}{6}
-\frac{1}{2}\frac{x_{1}^{2}+x_{2}^{2}}{x_{1}-x_{2}}\right]\mathrm{ln}\frac{x_{1}}{x_{2}}\nonumber \\
 & +\frac{x_{1}-1}{6}f\left(x_{1},x_{1}\right)+\frac{x_{2}-1}{6}f\left(x_{2},x_{2}\right)
 +\left[\frac{1}{3}-\frac{x_{1}+x_{2}}{6}-\frac{\left(x_{1}-x_{2}\right)^{2}}{6}\right]f
 \left(x_{1},x_{2}\right)\Bigg\}\label{pu}
\end{align}
in which $N_{C}=3\left(1\right)$ is the color degree of freedom of quarks
(leptons),
\[
F\left(x_{1},x_{2}\right)=\frac{x_{1}+x_{2}}{2}-\frac{x_{1}x_{2}}{x_{1}
-x_{2}}\mathrm{ln}\frac{x_{1}}{x_{2}}
\]
\[
f\left(x_{1},x_{2}\right)=\Bigg\{\begin{array}{c}
-2\sqrt{\Delta}\left[\mathrm{arctan}\frac{x_{1}-x_{2}+1}{\sqrt{\Delta}}
-\mathrm{arctan}\frac{x_{1}-x_{2}-1}{\sqrt{\Delta}}\right]\\
0\qquad\qquad\qquad\qquad\qquad\qquad\qquad\\
\sqrt{-\Delta}\,\mathrm{ln}\frac{x_{1}+x_{2}-1
+\sqrt{-\Delta}}{x_{1}+x_{2}-1-\sqrt{-\Delta}}\qquad\qquad\qquad\qquad\qquad
\end{array}\begin{array}{c}
\left(\Delta>0\right)\\
\left(\Delta=0\right)\\
\left(\Delta<0\right)
\end{array}
\]
with $x_{i}=m_{i}^{2}/M_{Z}^{2}$, and $\Delta=2\left(x_{1}+x_{2}\right)-\left(x_{1}-x_{2}\right)^{2}-1$,

It is important to indicate that if we consider new generations with non-SM hypercharges, the calculation show that it is 
possible to include them without conflicting with the allowed intervals for S and T\cite{ALVES}  for a wide range of masses.


\begin{thebibliography}{0}
\bibitem{SM1} S.L. Glashow, {\it Nucl. Phys.}, {\bf 22}, 579 (1961).
\bibitem{SM2} S. Weinberg, {\it Phys. Rev. Lett.}, {\bf 19}, 1264 (1967).
\bibitem{SM3} A. Salam, in: {\it  Elementary Particle Theory, Nobel Symposium}, 367 (Almqvist and Witsell,
Stockholm, 1968).
\bibitem{Z3} T. Riemann and J. Blumlein, {\it Proc. of the Zeuthern Workshop on Elementary Particles} {\it Nucl. Phys. B (Proc. Suppl.)}
{\bf B37} (1994). 
\bibitem{QCD1} D. Gross and F. Wilezek, {\it Phys. Rev. Letters}, {\bf 26}, 1343 (1973).
\bibitem{QCD2} M.D. Politzer, {\it Phys. Rev. Letters}, {\bf 26}, 1346 (1973).
\bibitem{2HDM} G. C. Branco, P. M. Ferreira, L. Lavoura, M. N. Rebelo, M.
Sher, and J. P. Silva, {\it Phys. Rep.} {\bf 516}, 1 (2012).
\bibitem{FERRE} P. M. Ferreira, R. Guedes, M. O. P. Sampaio, and R. Santos,
{\it JHEP} {\bf 12}, 067 (2014).
\bibitem{Higgs4f} Dipankar Das {\it et al.},  {\it Phys. Rev. D}, {\bf 97},011701(R)
 (2018).
\bibitem{SIN} Sin Kyu Kang {\it et al.} {\it Phys. Rev. D}, {\bf 98},095025 (2018). 
\bibitem{PESKIN} M.E. Peskin and T. Takeuchi, {\it Phys. Rev. D}, {\bf 46}, 381 (1992). 
\bibitem{PDG} M. Tanabashi {\it et al.}, (Particle Data Group), {\it Phys. Rev. D}, {\bf 98}, 030001 (2018). 
\bibitem{STU}V.A. Novikov {\it et al.}, {\it Physics Letters B} {\bf 529}, 111 (2002).
\bibitem{TAKEO} Takeo Inami {\it et al.} {\it Mod. Phys. Lett. A}, {\bf 10}, 1471 (1995).
\bibitem{Extra}M.\, Maltoni {\it et al.}, {\it Physics Letters B} {\bf 476}, 107 (2000).
\bibitem{ADRIANO} A.\,A. \,  Natale and P.\, S. Rodrigues da Silva {\it Mod. Phys. Lett. A}, {\bf 10}, 1829 (1994).
\bibitem{HE} Hong-Jian He {\it et al.}, {\it Phys. Rev. D}, {\bf 64} , 053004 (2001).
\bibitem{newLHC} Tobias Golling {\it Progress in Particle and Nuclear Physics} {\bf 90}, 156 (2016).
\bibitem{ALVES} Alexandre Alves  {\it et al.}, {\it JHEP}, {\bf 07}, 129  (2013).
\bibitem{BIZOT} Nicolas Bizot and Michele Frigerios, {\it JHEP}, {\bf 01}, 036  (2016).
\bibitem{FRAMPTON} P.\, H.\, Frampton {\it et al.} {\it Phys. Rep.}, {\bf 330}, 263 (2000).
\bibitem{CEBOLLA} Luís M. Cebola, D.Emmanuel-Costa, R. González Felipe, and C. Sim\~{o}es, 
{\it Phys. Rev. D}, {\bf 90}, 125037 (2014).
\bibitem{SIMOES} C. Sim\~{o}es, {\it Journal of Physics: Conference Series} {\bf 631}, 012084 (2015). 
\bibitem{ATLASalpha} Aaboud, M. {\it et al.}, {\it Eur. Phys Jour. C}, {\bf 77}, 872 (2017).
\bibitem{Becciolini} Diego  Becciolini {\it et al.}, {\it Phys. Rev. D}, {\bf 91},  015010 (2015). 
\bibitem{Llorente} Javier Llorente, Benjamin P. Nachman, {\it Nucl. Phys. B}, {\bf 936}, 106 (2018).
\bibitem{MoeDal} B. Acharya {\it et al.} {\it Int. J. Mod. Phys. A}, {\bf 29}, 1430050 (2014).

\end{thebibliography}
\end{document}